\documentclass{emulateapj}

\shorttitle{Einstein Cross}
\shortauthors{Bolton et al.\ }
 
\begin{document}
 
\def\lya{Ly$\alpha$}
\def\cf{{cf.,~}}
\def\ie{{i.e.,~}}
\def\eg{{e.g.,~}}
\def\etal{{et al.}}
\def\einstein{J1011$+$0143}
 
\title{A New Einstein Cross: A Highly Magnified, \\
Intrinsically Faint Lyman-Alpha Emitter at $z=2.7$}
 
\author{Adam S. Bolton\altaffilmark{1}}
\author{Leonidas A. Moustakas\altaffilmark{2}}
\author{Daniel Stern\altaffilmark{2}}
\author{Scott Burles\altaffilmark{3}}
\author{Arjun Dey\altaffilmark{4}}
\author{Hyron Spinrad\altaffilmark{5}}
 
\altaffiltext{1}{Harvard-Smithsonian Center for Astrophysics, 60 Garden
St., Cambridge, MA 02138 ({\tt abolton@cfa.harvard.edu})}
\altaffiltext{2}{Jet Propulsion Laboratory, California Institute of
Technology, 4800 Oak Grove Drive, Pasadena, CA 91109
({\tt leonidas@jpl.nasa.gov, stern@thisvi.jpl.nasa.gov})}
\altaffiltext{3}{Department of Physics and Kavli Institute for
Astrophysics and Space Research, Massachusetts Institute of Technology,
77 Massachusetts Avenue, Cambridge, MA 02139 ({\tt burles@mit.edu})}
\altaffiltext{4}{National Optical Astronomy Observatory, 950 N. Cherry
Ave., Tucson, AZ 85719 ({\tt dey@noao.edu})}
\altaffiltext{5}{Department of Astronomy, UC-Berkeley, Berkeley, CA
94720 ({\tt spinrad@astron.berkeley.edu})}

\submitted{ApJ Letters, in press}

\begin{abstract}

We report the discovery of a new Einstein cross at redshift $z_S =
2.701$ based on \lya\, emission in a cruciform configuration around an
SDSS luminous red galaxy ($z_L = 0.331$).  The system was targeted as a
possible lens based on an anomalous emission line in the
SDSS spectrum.  Imaging and spectroscopy from the W.M.~Keck Observatory
confirm the lensing nature of this system.  This is one of the
widest-separation galaxy-scale lenses known, with an Einstein radius
$\theta_{\rm E}\simeq 1.84$\,arcsec.  We present simple
gravitational lens models for the system and compute the intrinsic
properties of the lensed galaxy.  The total mass of
the lensing galaxy within the $8.8 \pm 0.1$ kpc enclosed by the lensed
images is $(5.2 \pm 0.1) \times 10^{11} M_{\sun}$.  The lensed galaxy is
a low mass galaxy (0.2$L_*$) with a high equivalent-width
Ly$\alpha$ line ($EW_{{\rm Ly}\alpha}^{\rm
rest} = 46 \pm 5$ \AA).  Follow-up studies of this lens system
can probe the mass structure of the lensing galaxy, and can provide a
unique view of an intrinsically faint, high-redshift, star-forming
galaxy at high signal-to-noise ratio.

\end{abstract}


\section{Introduction}

Strong gravitational lensing is a powerful tool for the measurement of
lensing galaxy masses and for the detailed study of magnified high-redshift
sources.  Multiple lensed images can directly constrain models for the
distribution of mass in the lens on the scale of the Einstein radius
$\theta_{\rm E}$.  In well-studied cases, these observations directly
test theories for the central dark-matter profile in both galaxies
\citep[e.g.,][]{kt03, dye_warren_05} and galaxy clusters (\eg
\citealt{kneib_2218_95, kneib_2218_96}; \citealt*{abdelsalam_98a,
abdelsalam_98b}).  Simultaneously, lensed background galaxies can be
magnified by factors of up to several tens, providing data of a quality
which would not otherwise be possible.  Such data have been used to
constrain the faint end of the high-redshift galaxy luminosity function
\citep[e.g.,][]{santos_2004}, study distant galaxies at wavelengths
which would be impractical without magnification
\citep*[e.g.,][]{chary_2005}, and probe detailed properties of
high-redshift galaxies
\citep{pettini_cb58}.

Systematic searches for new galaxy-scale strong gravitational lenses
are traditionally based on imaging detections
\citep[e.g.,][]{maoz_snapshot, gregg_2000, wisotzki_2002, inada_0924,
morgan_ctq327, richards_snap, pindor_2004}, although a handful of
lenses have been identified
in spectroscopic observations \citep{huch85, warren_0047_96, john03}.
The Sloan Lens ACS (SLACS) survey
(\citealt{slacs1, slacs2, slacs3}) efficiently
identifies new gravitational lenses by searching the spectroscopic
database of the Sloan Digital Sky Survey (SDSS; \citealt{york_sdss})
for systems consisting of low-redshift ($z \sim 0.1$--$0.4$) luminous
elliptical galaxies superposed with moderate-redshift ($z \sim
0.3$--$1$) star-forming galaxies (see also \citealt{bolton_speclens},
\citealt*{whw_2005}, and \citealt{whwdm_2006}).
This technique relies upon the detection of
{\em multiple} anomalous nebular emission lines in the SDSS fiber
spectra of lensing early-type galaxies.  In principle, the anomalous
emission line technique can also identify lenses with higher redshift,
\lya-emitting source galaxies \citep{warren_0047_96, h00, wil00}.  In
practice this approach has been less productive because strong,
high-redshift \lya\, emitters
are less numerous on the sky than moderate-redshift galaxies showing
oxygen and Balmer emission lines
to the SDSS line-flux limits, and the increase in
lensing cross section with source redshift does not overcome this
effect.

This {\em Letter} reports the discovery of a new
spectroscopically-selected strong gravitational lens with a
\lya-emitting galaxy as its source, SDSS~J101129.49$+$014323.3
(hereafter \einstein).  The lensed galaxy forms a highly
symmetric Einstein
cross which, by virtue of its large physical scale, provides leverage
for determining the dark-matter halo mass of the lensing elliptical.
The system also provides a highly magnified view of an intrinsically faint,
compact, and high-redshift star-forming galaxy.

For all calculations, we assume a universe with
$(\Omega_{\mathrm{M}},\Omega_{\Lambda},h) = (0.3,0.7,0.7)$.

\section{Discovery and Confirmation}

\begin{figure}
\centerline{\scalebox{0.48}{\includegraphics{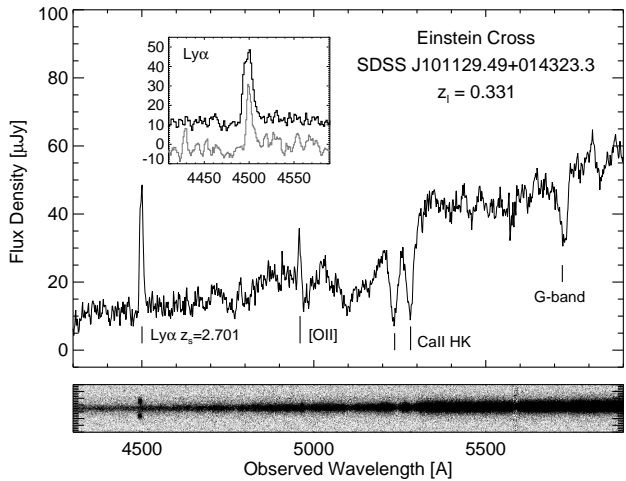}}}
\figcaption{\label{fig:spec}Spectrum of \einstein\ (one-dimensional
and two-dimensional forms), obtained with
Keck/LRIS on UT 2004 Nov 10, with features indicated.  The spectrum was
obtained through a 1\farcs5 slit aligned with the minor axis of
the lensing galaxy.  The 1D spectrum has been scaled to
match the spectrophotometry of the lensing elliptical.  \lya\
detail window also shows the SDSS line detection (grey), smoothed
by 5 pixels and shifted downwards by 20 $\mu$Jy.  Note the
apppearance of the lensed \lya\ line on either side of the
lensing galaxy continuum in the 2D spectrum.}
\end{figure}

\einstein\, was discovered on the basis of an anomalous emission line
in its SDSS spectrum at a vacuum wavelength of 4500~\AA\ (see
Fig.~\ref{fig:spec}).  Table~\ref{sdsstab} provides parameters for the
system as determined from the SDSS data.  The system was one of several
single-line lens candidates selected from the luminous red galaxy
(LRG) sample of the SDSS \citep{eisenstein_lrg}.  These candidates were
identified by subtracting best-fit galaxy spectrum templates from the
SDSS spectra of the LRGs, and then
searching for significant ($\geq 7 \sigma$) residual
emission.  The search was done via convolution with a Gaussian kernel
matched to the SDSS spectral resolution ($\approx$150 km\,s$^{-1}$).
Spectrum noise estimates were empirically rescaled in a
manner similar to that of \citet{bolton_speclens}.  Our search was
confined to wavelengths short-ward of 6500~\AA\, since at redder
wavelengths one must contend with a deluge of low-redshift H$\alpha$
detections.  Candidates were subjected to several pruning
steps to reject (1) unmodeled LRG rest-frame absorption/emission,
(2) night-sky emission residuals, (3) spectra of very poor data quality,
(4) line profiles narrower than the SDSS spectral resolution, and
(5) lines other than \lya\ as judged by multiple
emission lines at the same redshift.  In preparation for a
single observing run, 21 priority candidates were
identified in the ranges 19h--01h and 05h--11h.  SDSS imaging of
these candidates was inspected to verify the
absence of bright neighboring galaxies
that could account for anomalous emission lines.

\begin{table} 
\begin{center} 
\caption{\label{sdsstab} SDSS parameters for \einstein}
\begin{tabular}{ccc} \hline\hline \multicolumn{1}{c}{SDSS ID$^{a}$} &
\multicolumn{1}{c}{J2000 RA$^b$} & \multicolumn{1}{c}{J2000 Dec.$^b$ } \\ \hline
502-51957-172 & 10h11m29.50s & $+$01d43m23.4s \\ \hline \\ \end{tabular}

\begin{tabular}{ccccc} 
\hline
\hline 
\multicolumn{1}{c}{$g,r,i$ [AB]$^{c}$}
& \multicolumn{1}{c}{$R_{\rm eff}$ [$''$]$^{d}$} &
\multicolumn{1}{c}{$z_{\rm lens}$} & \multicolumn{1}{c}{$z_{\rm
source}$} & \multicolumn{1}{c}{$\sigma_v$ [km\,s$^{-1}$]$^{e}$} \\ \hline
$19.4,18.1,17.5$ & $1.71\pm0.1$ & 0.3308 & 2.701 & $259\pm16$ \\
\hline \\ 
\end{tabular} 
\end{center}
\smallskip
$a$: SDSS PLATE-MJD-FIBERID (MJD $=$ modified Julian date)\\
$b$: Coordinates accurate to $0\farcs 1$ \\
$c$: Extinction-corrected DeVaucouleurs model magnitudes. \\
$d$: Effective radius at intermediate axis. \\ 
$e$: Stellar velocity dispersion as measured from SDSS spectroscopy. \\
\end{table}

In order to confirm the SDSS anomalous emission line detection, initial
follow-up of \einstein\, was conducted on the night of UT 2004 November
9 using the Low Resolution Imaging Spectrometer (LRIS;
\citealt{oke_lris}) with the Keck I Telescope.  A large slit of
1$\farcs$5 width was used to maximize the aperture with which to detect
any lensed emission-line flux.  The slit, which is still smaller than
the 3\arcsec diameter fibers used in the discovery SDSS spectroscopy,
was aligned with the parallactic angle.  The exposure time was 10~min
and conditions were photometric with 0\farcs8 seeing.  The
two-dimensional spectrum of \einstein\, showed extended emission-line
morphology at the wavelength of the SDSS detection (4500~\AA).  Three
other candidate systems observed similarly on the same night showed no
evidence of the line emission detected by SDSS, and hence are presumed
associated with noise or cosmic rays in the SDSS
spectroscopy\footnotemark.\footnotetext{The three non-detections are
identified by the following SDSS PLATE-MJD-FIBERID values, with
J2000 right ascensions and declinations and anomalous line
wavelengths given parenthetically:
385-51877-437 (23h39m24.33s $+$00d32m34.2s, 5116 \AA);
389-51795-373 (00h12m49.09s $+$00d53m21.9s, 6180 \AA); and
656-52148-411 (00h44m52.11s $-$09d04m54.6s, 4302 \AA).}
Based on the confirmation of the SDSS line emission, \einstein\ was
observed again on the night of UT 2004 November 10 with LRIS $B$-band
(600~s) and $I$-band (420~s) imaging, as well as further, deeper
(30~min) spectroscopy.  The $B$-band image shows a distinctive
quadruple-image cross morphology, aligned with the principal axes of
the LRG and characteristic of strong gravitational lensing
(Fig.~\ref{fig:imfig}).  For the additional spectroscopy, the
1$\farcs$5 slit was aligned with the minor axis of the LRG.  The
resulting spectrum is shown in Fig.~\ref{fig:spec}.

Although based on a single emission line, the identification of the
cross emission line at 4500~\AA\, with redshifted \lya\ is
secure for the following reasons.  First, the Keck detections confirm
that the emission line is real and not an artifact or night-sky
residual.  Second, the spectroscopic data (particularly the
higher resolution SDSS
detection) show the classic, asymmetric, self-absorbed morphology
typical of \lya\, emission.  Third, the wavelength of the line does not
correspond to any common emission wavelength at the redshift of the
LRG.  Finally, the only other possible common emission lines observable
at 4500~\AA\, are [\ion{O}{2}] $\lambda 3727$ or Balmer lines of order
H$\gamma$ and higher.  The [\ion{O}{2}] identification is strongly
disfavored by both the line morphology and the absence of corresponding
[\ion{O}{3}] and H$\alpha$ emission.  A Balmer-series identification is
likewise disfavored by the absence of other Balmer-series lines.

\section{Image Analysis and Lens Modeling}

\begin{figure}
\plotone{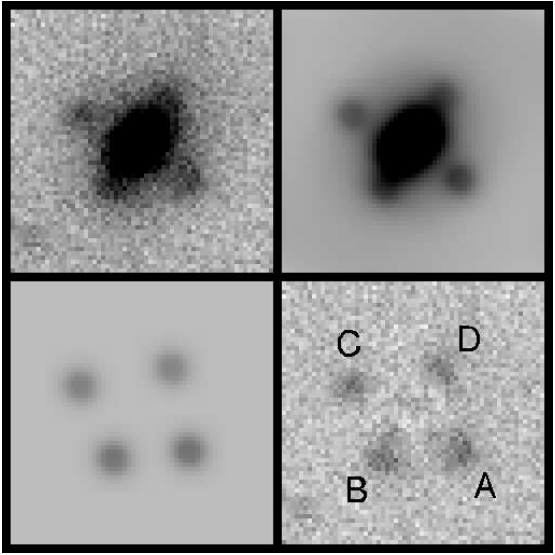}
\figcaption{\label{fig:imfig}$B$-band data, model, and residual images
of \einstein.  Shown from upper left to lower right are direct image,
composite model, lensed-image model and galaxy model-subtracted
residual.  Image region is $8\farcs 2 \times 8\farcs 2$, with north up
and east to the left.} 
\end{figure}

The four symmetrically-arranged lensed images are well-detected in the
Keck $B$-band image (Fig.~\ref{fig:imfig}).  Even with a $B$-band
seeing of 1\arcsec, we can successfully fit simple analytic models
given the wide image separation.  We measure relative positions of the
lensed images with respect to the lensing galaxy by fitting a model
composed of four PSFs (for the four lensed images) and a radial
b-spline with a quadrupole angular dependence (for the lens galaxy; see
\citealt{slacs1} for a full discussion of radial b-splines).
Table~\ref{impars} gives the parameters of the derived image
components. We fit over an image cutout of $61 \times 61 = 6721$
pixels, subtending a region $8\farcs2$ on a side.  The composite image
model has a total of 38 free parameters, and the $\chi^2$ per degree of
freedom of the fit is $\chi_R^2 = 1.21$.  Fig.~\ref{fig:imfig} shows
data, model, and residual images of the fitted region about the lens.
The model implies a flux ratio of 10.77 between the total $B$-band flux
of the lensing galaxy and the summed flux of the four lensed images.

\begin{table}
\begin{center}
\caption{\label{impars}Measured parameters of $B$-band image
components}
\begin{tabular}{cccc}
\hline\hline
\multicolumn{1}{c}{} &
\multicolumn{1}{c}{$\Delta\alpha$} &
\multicolumn{1}{c}{$\Delta\delta$} &
\multicolumn{1}{c}{Magnitude} \\
\multicolumn{1}{c}{Image} &
\multicolumn{1}{c}{($''$E)$^a$} &
\multicolumn{1}{c}{($''$N)$^a$} &
\multicolumn{1}{c}{($B_{\rm AB}$)$^b$} \\
\hline
G & $\equiv$ 0 & $\equiv$ 0 & $19.86 \pm 0.01$ \\
A & $-1.55\pm0.02$ & $-1.14\pm0.02$ & $23.79\pm0.04$\\
B & $+0.81\pm0.03$ & $-1.36\pm0.03$ & $23.84\pm0.04$\\
C & $+1.86\pm0.03$ & $+0.91\pm0.03$ & $24.05\pm0.05$\\
D & $-1.00\pm0.03$ & $+1.46\pm0.04$ & $24.10\pm0.05$\\
\hline \\
\end{tabular}
\end{center}
\smallskip
$a$: Positions relative to Table~\ref{sdsstab} coordinates.\\
$b$: Quoted uncertainties do not include overall magnitude zeropoint
uncertainty of $\approx 0.1$.\\
\end{table}

The lensed images are not readily visible in the $I$-band Keck image.
This is likely due to a combination of the intrinsic faintness of the lensed
images at this wavelength and the red color of the lensing elliptical.
In addition, b-spline galaxy subtraction of the $I$-band image is
complicated by an asymmetric PSF.  For these reasons, no quantitative
analysis of the $I$-band image is presented.

Given the limited spatial resolution and
signal-to-noise ratio of our current data,
we consider only a standard singular isothermal ellipsoid (SIE) lens
model and a singular isothermal sphere with external shear
(SIS$\gamma$).  Both models are simple, analytic, physically motivated
and free from severe parameter degeneracies.  The angular Einstein
radius $b$ of the SIS model is related to the velocity dispersion
$\sigma$ of the lensing distribution through
$b_{\mathrm{SIS}} = 4 \pi (\sigma^2/c^2)
(D_{\mathrm{LS}} / D_{\mathrm{S}})$, where
$D_{\mathrm{LS}}$ and $D_{\mathrm{S}}$ are angular-diameter distances
from lens to source and observer to source.  To assign a lensing
velocity dispersion to an SIE model with
minor-to-major axis ratio $q < 1$,
we adopt the intermediate-axis
normalization of \citet*{kormann_sie}, whereby the mass interior to a
given iso-density contour at fixed $b$ is constant with changing $q$.

The data provide 12 constraints (RA, Dec., and flux for each of the
four lensed images).  We implement both SIE and SIS$\gamma$ models with
a total of 8 free parameters.  The best-fit (minimum-$\chi^2$) values
for these parameters are determined using custom IDL routines that
invoke the MPFIT implementation of the Levenberg-Marquardt minimization
algorithm \citep{more_wright}.
Table~\ref{lenspars} gives fitted values of the parameters
for both models.  Although the alignment of the four lensed images with
the principal axes of the LRG suggest that the intrinsic
ellipticity of the lens galaxy is responsible for the
quadrupole of the lensing potential,
the SIS$\gamma$ model provides a better fit than
the SIE to both the image positions and fluxes.  We note that using
image fluxes as lens-model constraints is a legitimate tactic, given
that the source is likely to be a small but extended high-redshift
star-forming galaxy, and thus immune to both the microlensing
perturbations and temporal variability that affect lensed quasar image
fluxes.  Large-mass substructures in the lens could, however, perturb the
image fluxes from the predictions of smooth models.

\begin{table}
\begin{center}
\caption{\label{lenspars}Lens model parameters}
\begin{tabular}{lcc}
\hline \hline
\multicolumn{1}{c}{Model} & 
\multicolumn{1}{c}{SIE}   &
\multicolumn{1}{c}{SIS}   \\
\multicolumn{1}{c}{parameter} & 
\multicolumn{1}{c}{no shear}   &
\multicolumn{1}{c}{plus shear} \\
\hline
$b$ (arcsec) & $1.87 \pm 0.02$ & $1.84 \pm 0.02$ \\
$b$ (kpc) & $8.9 \pm 0.1$ & $8.8 \pm 0.01$ \\
Quadrupole & $q = 0.77 \pm 0.03$ & $\gamma = 0.078 \pm 0.008$ \\
PA ($^{\circ}$E of N) & $148.4 \pm 0.1$ & $148.4 \pm 0.1$ \\
Rel.\ Lens RA      (\arcsec E)$^a$ & $-0.03 \pm 0.02$ & $-0.01 \pm 0.02$ \\
Rel.\ Lens Dec.\   (\arcsec N)$^a$ & $+0.10 \pm 0.02$ & $+0.09 \pm 0.02$ \\
Rel.\ Source RA    (\arcsec E)$^a$ & $+0.01 \pm 0.02$ & $+0.02 \pm 0.02$ \\
Rel.\ Source Dec.\ (\arcsec N)$^a$ & $+0.07 \pm 0.02$ & $+0.06 \pm 0.02$ \\
Source Magnitude ($B_{\rm AB}$)$^b$ & $25.9 \pm 0.1$ & $26.0 \pm 0.1$ \\
\hline
Total Magnif.\ & $23 \pm 3$ & $26 \pm 2$ \\
$\chi^2$ (d.o.f.) & $61.1 (4)$ & $23.2 (4)$ \\
$\chi^2_{\mbox{posn}}$, $\chi^2_{\mbox{flux}}$ &
35.4, 25.7 & 10.9, 12.3 \\
\hline
Lens-model vdisp.\ (km\,s$^{-1}$) & $288 \pm 2$ & $285 \pm 2$ \\
Mass within $b$ ($10^{11} M_{\sun}$) & $5.4 \pm 0.1$ & $5.2 \pm 0.1$ \\
\hline
\end{tabular}
\end{center}
\smallskip
$a$:  Positions relative to Table~\ref{sdsstab} coordinates.\\
$b$:  Best-fit lens model parameter value for unlensed
 point-source magnitude. \\
\end{table}

Adopting $b=1.84 \pm 0.02$ from the best-fit SIS$\gamma$
model, the mass enclosed by the lensed images is
$(5.2 \pm 0.1) \times 10^{11} M_{\sun}$.
The corresponding lensing velocity dispersion
is $\sigma_{\mathrm{SIS}} = 285 \pm 2$ km\,s$^{-1}$.
This is larger
than the stellar value from the SDSS spectrum of $\sigma_v = 259 \pm
16$ km\,s$^{-1}$ (based on a median SNR per pixel of 10 and a
resolution of $\approx$ 150 km\,s$^{-1}$), although
the significance of the difference is less than two standard
deviations.  The Keck spectroscopic data do not permit an
independent determination of the stellar velocity dispersion
due to their lower resolution ($\approx$ 600 km\,s$^{-1}$).

\section{Properties of the Lensed Galaxy}

The \lya\ flux in each of the cross components observed on UT 2004
November 10 is $2.8 \times 10^{-16}\, {\mathrm{erg}}\,
{\mathrm{cm}}^{-2} {\mathrm{s}}^{-1}$.  For all four lensed components
having roughly equal \lya\ flux, the implied total \lya\ flux for the
Einstein cross is $1.1 \times 10^{-15}\, {\mathrm{erg}}\,
{\mathrm{cm}}^{-2} {\mathrm{s}}^{-1}$.  For our assumed cosmology and
the magnification of the best-fit SIS$\gamma$ model, we derive an
intrinsic Ly$\alpha$ luminosity of $2 \times 10^{42}\, {\mathrm{erg}}\,
{\mathrm{s}}^{-1}$.  \citet{dawson_lya} present the most recent,
comprehensive derivation of the luminosity function of high-redshift
\lya\, emitters.  They conclude that $L_* \approx 10 \times 10^{42}\,
{\mathrm{erg}}\, {\mathrm{s}}^{-1}$, with no evidence for evolution in
the \lya\, luminosity function between $z \approx 3$ and $z \approx
6$.  This implies that \einstein\, is an intrinsically faint \lya\,
emitter, with a luminosity $\approx 0.2 L_*$, but with a magnification
that makes it among the brightest high-redshift \lya\, emitters known.

The current Keck spectroscopy is of
insufficient depth to allow detection of the faint continuum of the
lensed galaxy.  We thus combine SDSS photometry (Table~\ref{sdsstab},
assuming statistical errors of 0.1 mag) and Keck imaging
and spectroscopy to estimate the broadband magnitude and \lya\,
equivalent width of the lensed galaxy.  We assume the shape of the lens
galaxy continuum to be given by the best-fit template spectrum from the
Princeton SDSS redshift pipeline\footnote{\tt
http://spectro.princeton.edu}.  For the lensed galaxy, we assume a
simple model for the continuum shape that includes \lya\, forest
absorption:
\begin{equation} 
f_{\nu} (\lambda) = \left\{
\begin{array}{ll}
  f_{\rm cont},     & \lambda/(1+z_s) > 1216 {\rm \AA\ } ; \\            
  0.55 f_{\rm cont} & 1026 < \lambda/(1+z_s) \le 1216 {\rm \AA\ }.
\end{array}
\right.  
\end{equation}
This model is a reasonable approximation to the
composite spectrum of the strongest \lya -emitting LBGs at similar
redshifts published by \citet{shapley_lbg}.  We integrate these
continuum models over $g$ and $B$ filter curves, together with the
known contribution from the measured \lya\, line flux.  We then deduce
the relative normalization of each continuum component from (1) the
measured SDSS $g$ band flux for the entire system, and (2) the measured
Keck $B$-band flux ratio of 10.77 of the lens to the source (continuum
plus line, summed over all four lensed images).  The results give
$B_{\rm AB,lens} = 19.9 \pm 0.1$ and $B_{\rm AB,source} = 22.5 \pm 0.1$.
Taking the total magnification of the best-fit SIS$\gamma$ model, we
find an unlensed magnitude for the source galaxy of $B_{\rm
AB,source}^{\rm unlensed} = 26.0 \pm 0.1$.  The implied continuum flux
density of the lensed galaxy red-ward of \lya\, (including the observed
magnification) is $4.3\pm0.4$~$\mu$Jy, giving an observed equivalent
with of $EW_{{\rm Ly}\alpha}^{\rm obs} = 170 \pm 20$ \AA.  Correcting
for cosmological expansion, this becomes $EW_{{\rm Ly}\alpha}^{\rm
rest} = 46 \pm 5$ \AA, a value typical of confirmed high-redshift \lya\
emitters \citep[e.g.,][]{dawson_lya}.

\section{Summary and Conclusions}

We present the discovery of a new high-redshift Einstein-cross
gravitational lens, \einstein.  The lens galaxy is a bright elliptical
at $z_{\rm lens} = 0.331$, while the lensed source is a $\sim 0.2 L_*$
\lya -bright, star-forming galaxy at redshift $z_{\rm source} = 2.701$.
Though such high-$z$ \lya\ lenses appear to be
much less numerous in the SDSS than
lenses with lower redshift ($z \la 1$) oxygen- and Balmer-line
emitting sources
\citep{bolton_speclens, whw_2005, slacs1, whwdm_2006}, the discovery of
\einstein\ demonstrates their presence.  Depending on the luminosity
function slope of the \lya -emitting source population, deeper
spectroscopic surveys could yield an appreciable sample of such lens systems.

As with all gravitational lenses, this system offers a powerful tool
for measuring the mass in the lensing galaxy.  \einstein\ is of
particular interest for its relatively wide ($\sim 4\arcsec$) image
separation, probing the lens galaxy at a radius within which the
contributions of luminous and dark matter are expected to be
comparable.  As with the sample of lenses discovered by the SLACS
survey \citep{slacs1}, the image of the lens galaxy is not overwhelmed
by lensed-quasar images and is thus accessible to accurate
photometric and dynamical observations.
The high degree of symmetry of the
image configuration suggests that a largely model-independent test of
the relative degree of flattening between the mass and light
distributions in the early-type lens galaxy will be enabled by
high-resolution imaging.
A modest Cycle~15 {\it HST}/ACS program to image this system has
been awarded two orbits (P.I. Moustakas).

\einstein\ also offers a highly magnified view of a sub-$L_*$ starforming
galaxy at high redshift.  This discovery is thus complementary to the
more luminous Lyman-break galaxy MS1512$-$cB15 \citep{yee_cb58}, which
is also strongly magnified by gravitational lensing
\citep{seitz_cb58}.  The magnification of \einstein\ suggests this
system as a target for deep spectroscopic studies of the high-redshift
IGM that would otherwise be infeasible due to the intrinsic source
faintness.

\acknowledgments Some of the data presented herein were obtained at the
W.M.~Keck Observatory, which is operated as a scientific partnership
among the California Institute of Technology, the University of
California and the National Aeronautics and Space Administration. The
Observatory was made possible by the generous financial support of the
W.M.~Keck Foundation.  The authors wish to recognize and acknowledge
the very significant cultural role and reverence that the summit of
Mauna Kea has always had within the indigenous Hawaiian community; we
are most fortunate to have the opportunity to conduct observations from
this mountain.  The work of LAM and DS was carried out at Jet
Propulsion Laboratory, California Institute of Technology, under a
contract with NASA.  AD acknowledges support from NOAO, which is
operated by the Association of Universities for Research in Astronomy
(AURA), Inc.\ under a cooperative agreement with the
National Science Foundation.

\end{document}